\documentclass[sigconf]{acmart}
\usepackage{enumitem}
\usepackage{cleveref}
\usepackage{booktabs}

\usepackage{graphicx}
\usepackage{float}
\usepackage{subcaption}
\usepackage{hyperref}
\usepackage{multirow}

\crefname{table}{Table}{Tables}  
\crefname{figure}{Figure}{Figures}
\crefname{equation}{Equation}{Equations}

\AtBeginDocument{%
  }

\setcopyright{acmlicensed}
\copyrightyear{2018}
\acmYear{2018}
\acmDOI{XXXXXXX.XXXXXXX}
\acmConference[Conference acronym 'XX]{Make sure to enter the correct
  conference title from your rights confirmation email}{June 03--05,
  2018}{Woodstock, NY}
\acmISBN{978-1-4503-XXXX-X/2018/06}

\begin{document}

\title{GAP-Net: Calibrating User Intent via Gated Adaptive Progressive Learning for CTR Prediction}

\author{Shenqiang Ke}
\affiliation{%
  \institution{Meituan}
  \city{Beijing}
  \country{China}}
\email{keshenqiang@meituan.com}

\author{Jianxiong Wei}
\affiliation{%
  \institution{Meituan}
  \city{Beijing}
  \country{China}}
\email{weijianxiong@meituan.com}

\author{Qingsong Hua}
\affiliation{%
  \institution{Meituan}
  \city{Beijing}
  \country{China}}
\email{huaqingsong@meituan.com}

\renewcommand{\shortauthors}{Ke et al.}

\begin{abstract}
Sequential user behavior modeling is pivotal for Click-Through Rate (CTR) prediction yet is hindered by three intrinsic bottlenecks: (1) the "Attention Sink" phenomenon, where standard Softmax compels the model to allocate probability mass to noisy behaviors; (2) the Static Query Assumption, which overlooks dynamic shifts in user intent driven by real-time contexts; and (3) Rigid View Aggregation, which fails to adaptively weight heterogeneous temporal signals according to the decision context. To bridge these gaps, we propose GAP-Net (Gated Adaptive Progressive Network), a unified framework establishing a "Triple Gating" architecture to progressively refine information from micro-level features to macro-level views. GAP-Net operates through three integrated mechanisms: (1) Adaptive Sparse-Gated Attention (ASGA) employs micro-level gating to enforce sparsity, effectively suppressing massive noise activations; (2) Gated Cascading Query Calibration (GCQC) dynamically aligns user intent by bridging real-time triggers and long-term memories via a meso-level cascading channel; and (3) Context-Gated Denoising Fusion (CGDF) performs macro-level modulation to orchestrate the aggregation of multi-view sequences. Extensive experiments on industrial datasets demonstrate that GAP-Net achieves substantial improvements over state-of-the-art baselines, exhibiting superior robustness against interaction noise and intent drift.

\end{abstract}






\maketitle

\section{Introduction}

Click-Through Rate (CTR) prediction stands as the cornerstone of modern online advertising~\cite{guo2017deepfm,ma2018entire} and recommendation systems~\cite{covington2016deep,kang2018self,xu2025climber}, playing a pivotal role in optimizing traffic allocation and maximizing platform revenue. In this domain, the dominant paradigm has transitioned from static user profiling to \textit{Sequential User Behavior Modeling}~\cite{kang2018self,li2023text}. By encoding variable-length interaction histories into expressive representations, sequential models capture complex, non-linear dependencies between historical behaviors and current candidates, providing a robust foundation for identifying fine-grained user preferences.

To effectively encode these behavioral signals, the research landscape has evolved from foundational attention mechanisms to high-capacity architectures. Early works like DIN~\cite{zhou2018deep} and DIEN~\cite{zhou2019deep} pioneered the use of Target Attention, enabling models to track the dynamic evolution of user interests relative to candidate items. Building on this foundation, efforts to enhance model expressiveness have expanded along two principal dimensions. The first dimension involves Extending Sequence Length, where retrieval-based frameworks (e.g., SIM~\cite{pi2020search}, ETA~\cite{chen2021end}) filter relevant actions from massive historical data, thereby broadening the receptive field from short-term sessions to lifelong horizons. The second dimension focuses on Broadening Information Width to enrich context representation. Beyond traditional attribute (e.g., DIF-SR~\cite{xie2022decoupled}), a burgeoning trend leverages the semantic capabilities of Large Language Models~\cite{yang2022chinese,hu2024minicpm} to integrate heterogeneous multi-modal side information—such as deep textual semantics and visual cues~\cite{zhang2025moon}—into the recommendation loop. Fundamentally, both streams share a unified objective: augmenting the model's input space with longer historical sequences and wider information channels.

\begin{figure*}[!t]
    \centering
    \includegraphics[width=1.0\linewidth, alt={Illustration of three intrinsic blind spots.}]{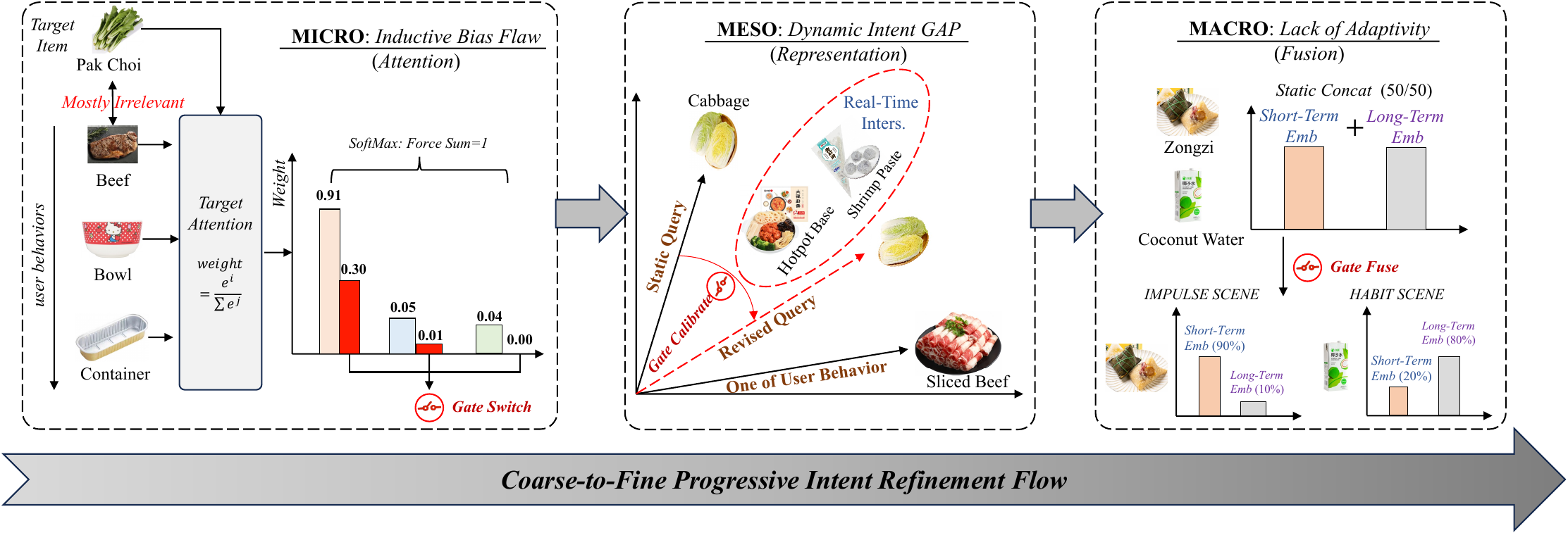}
    \caption{Illustration of intrinsic blind spots. (a) \textit{Micro-Level: Inductive Bias Flaw.} The standard Softmax enforces a strict sum-to-one constraint, forcing the model to assign spurious weights to irrelevant noise. (b) \textit{Meso-Level: Representation Gap.} Static target embeddings fail to capture dynamic intent shifts driven by real-time context cues (e.g., shifting from "daily meal" to "social dining"), leading to intent misalignment. (c) \textit{Macro-Level: Rigid Fusion.} Static aggregation lacks adaptivity, failing to dynamically modulate the trade-off between short-term impulses and long-term habits based on the specific decision scenario.}
    \vspace{-0.5cm}
    \label{fig:blind_spots}
\end{figure*}

However, this pursuit of "Longer and Wider" sequences masks intrinsic flaws in the Micro-level Interaction Mechanism. Most state-of-the-art models still rely on standard Softmax Attention as the atomic operation, which suffers from three critical blind spots:

(1) Lack of "Rejection" Capability: The standard Softmax function enforces a strict sum-to-one normalization, compelling the model to distribute its entire probability mass across the history sequence. This structural rigidity leads to the "Attention Sink" phenomenon, yet with a critical distinction in manifestation between NLP and RecSys. In LLMs, residual attention scores—when no meaningful token is found—are typically absorbed by specific "sink tokens" (e.g., the initial token or delimiters)~\cite{qiu2025gated}. In contrast, user behavior sequences generally lack such dedicated canonical sinks. Consequently, the "sink" in sequential recommendation inevitably manifests as noisy behaviors (e.g., accidental clicks or irrelevant history). The model is effectively forced to assign spurious relevance to these noise artifacts, amplifying noise accumulation across long sequences. This highlights an urgent need for sparsity-inducing mechanisms that enable true "zero-attention" to irrelevant features~\cite{ma2019hierarchicalgatingnetworkssequential,zaheer2021bigbirdtransformerslonger}.

(2) Static Interaction Paradigm: The conventional assumption that the Target Item (Query) serves as a static anchor overlooks the fluidity of user intent, which is inherently driven by real-time context. This creates a semantic gap, as the user's intention toward the same item can vary drastically depending on situational cues. For instance, consider the target item \textit{"Baby Cabbage"}. On a weekday, it typically signifies a routine \textit{"daily meal"} intent; however, on a weekend, if it appears alongside items like \textit{"Hot Pot Soup Base"} or \textit{"Shrimp Paste"}, the intent shifts significantly to a \textit{"social dining"} (hot pot) context. Existing models that treat the target embedding as immutable fail to capture this nuance, often retrieving historically relevant but contextually mismatched behaviors. While approaches like DIEN~\cite{zhou2019deep} model interest evolution, they typically fix the query vector during the retrieval process.

(3) Rigid View Aggregation: Existing multi-view frameworks typically resort to static concatenation or summation to merge heterogeneous temporal signals (e.g., real-time triggers vs. long-term memories). This context-agnostic strategy fails to dynamically modulate the importance of each view, allowing macro-level noise from irrelevant time windows to dilute valid signals—especially when user intent shifts rapidly between habitual and impulsive modes. Although methods like DIF-SR~\cite{xie2022decoupledinformationfusionsequential} and MIND~\cite{li2019multiinterestnetworkdynamicrouting} decouple short- and long-term interests, their fusion stages lack adaptivity. While recent advances propose context-gated fusion mechanisms~\cite{CHEN2025111785,li2025multimodalanchorgatedtransformer} or reinforcement learning-based selection~\cite{ji2025hierarchicalreinforcementlearningtemporal}, a unified adaptive framework remains elusive.

To remedy these fundamental defects, we draw inspiration from parallel advancements in Large Language Models (LLMs). Noting the success of Gated Attention in stabilizing massive activations, we rethink the systemic potential of gating in sequential recommendation. While gating units (e.g., GRU gates~\cite{guo2022evolutionary,chang2021sequential}, MoE routers~\cite{bian2023multi,xu2024mome}) have been sporadically applied in RecSys, they are predominantly utilized as isolated tools for sub-tasks like structure pruning or static expert routing. A critical gap remains: current approaches overlook gating as a \textit{comprehensive defense mechanism} against noise. This motivates our core inquiry: \textit{Can we establish a systematic gating philosophy that orchestrates sparsity-based denoising and context-aware calibration across all granularities?}

Addressing these challenges, we propose the \textbf{G}ated \textbf{A}daptive \textbf{P}rogressive Network (GAP-Net), a unified framework based on a "Triple Gating" philosophy. The core objective of GAP-Net is to systematically eliminate interaction noise and calibrate user intent through gating mechanisms orchestrated at the micro, meso, and macro levels. Specifically, the framework consists of three integrated modules: (1) Adaptive Sparse-Gated Attention (ASGA) operates at the \textit{micro-feature level}, introducing a learnable sparse gating mechanism that relaxes the strict sum-to-one constraint to suppress feature-level noise and mitigate the "Attention Sink" problem. (2) Gated Cascading Query Calibration (GCQC) operates at the \textit{meso-intent level}, abandoning the static query assumption to construct a cascaded channel that progressively refines the query vector using real-time contextual triggers, ensuring the retrieval process aligns with dynamic intent. (3) Context-Gated Denoising Fusion (CGDF) operates at the \textit{macro-view level}, utilizing a purified decision context to adaptively modulate the contribution of heterogeneous temporal views (e.g., real-time vs. long-term sequences), preventing noise propagation from irrelevant time windows.

The main contributions of this paper are summarized as follows:
\begin{itemize}[leftmargin=*, topsep=0pt, partopsep=0pt]
    \item We systematically identify three intrinsic bottlenecks in existing CTR models—the "Attention Sink" phenomenon, the "Static Intent" assumption, and "Rigid View Aggregation"—and introduce a unified "Triple Gating" philosophy to resolve them.
    \item We propose GAP-Net, a novel architecture that orchestrates \textit{micro-level} sparse attention (ASGA), \textit{meso-level} intent calibration (GCQC), and \textit{macro-level} dynamic fusion (CGDF) to ensure comprehensive noise resilience.
    \item Extensive experiments on industrial datasets demonstrate that GAP-Net achieves state-of-the-art performance, exhibiting superior robustness against interaction noise and intent drift.
\end{itemize}

\section{Related Work}

In this section, we review the evolution of Target-Attention-based models and gating mechanisms, highlighting the critical limitations in existing literature that motivate the design of GAP-Net.

\textit{Sequential Recommendation.} 
Sequential modeling serves as the backbone of CTR prediction, having evolved from early pooling methods to sophisticated attention-based paradigms. DIN~\cite{zhou2018deep} pioneered the Target Attention mechanism to capture diverse user interests relative to candidate items, while successors like DIEN~\cite{zhou2019deep} and DSIN~\cite{feng2019deep} further incorporated interest evolution and session-level dynamics. However, these foundational approaches predominantly rely on the standard Softmax function, which enforces a strict sum-to-one constraint. This design inherently assumes the presence of relevant items within any history window, compelling the model to allocate probability mass even to purely noisy or irrelevant behaviors. Consequently, these models lack the capability for "soft rejection," leading to the accumulation of spurious signals—a phenomenon typically referred to as the "Attention Sink."

To address scalability and expressiveness, recent research has bifurcated into two streams. Regarding sequence \textit{length}, search-based models like SIM~\cite{pi2020search} and ETA~\cite{chen2021end} introduced two-stage retrieval (GSU/ESU) to handle life-long sequences. Nevertheless, these methods rely on rigid hard-retrieval metrics (e.g., category matching) that act as static filters. They fail to account for the fluidity of user intent, often retrieving historically relevant but mismatched items (e.g., recommending "daily supplies" during a "gift-giving" context). While recent works such as TWIN~\cite{chang2023twin,si2024twin} and LONGER~\cite{chai2025longer} push boundaries to ultra-long horizons, they prioritize sequence length over interaction quality, leaving the underlying sensitivity to noise unresolved. Regarding sequence \textit{width}, methods such as DIF-SR~\cite{xie2022decoupled}, ASIF~\cite{wang2024aligned}, and CAIN~\cite{guo2025context} focus on fusing heterogeneous side information. However, they typically employ context-agnostic aggregation strategies (e.g., concatenation), which lack the adaptivity to down-weight irrelevant views when the dominant signal shifts between real-time triggers and long-term habits. In contrast, GAP-Net re-imagines the core interaction mechanism, introducing a unified multi-level gating philosophy to systematically calibrate user intent and dynamically fuse heterogeneous views.

\textit{Gated Mechanism in Recommendation System.} 
Gating mechanisms have transitioned from foundational RNN components into versatile tools for feature selection and dynamic routing. DCN V2~\cite{wang2021dcn} and AdaSparse~\cite{yang2022adasparse} integrated gating to induce adaptive structural sparsity, while PEPNet~\cite{chang2023pepnet} leverages GateNet for personalized embedding pruning. In multi-task settings, frameworks like PLE~\cite{tang2020progressive} and M2M~\cite{zhang2022leaving} employ gating to route samples to specific experts. Despite their success, these approaches typically deploy gating as isolated modules for structural pruning or static routing. They treat gating primarily as a static feature selector rather than an interaction refinement mechanism, failing to orchestrate gating dynamically across the temporal dimension. Consequently, they cannot effectively prevent noise propagation during the sequential evolution of user intent. GAP-Net fills this void by establishing a unified "Triple Gating" architecture that progressively filters noise and refines intent from micro-level features to macro-level views.

\section{Method}
\begin{figure*}
    \centering
    \includegraphics[width=\linewidth, alt={An Overview of the proposed GAP-Net}]{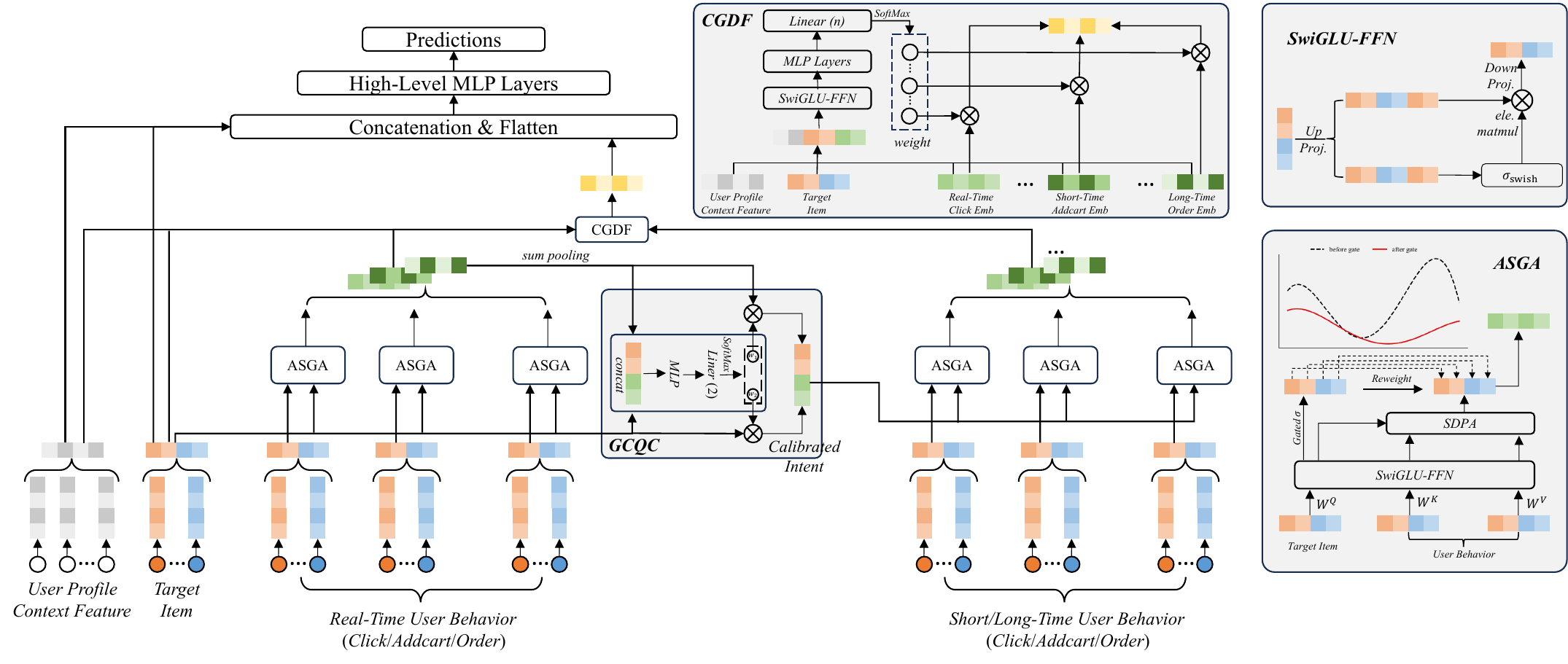}
    \caption{An Overview of the proposed GAP-Net. The framework employs a "Triple Gating" architecture for progressive denoising and calibration: (a) Micro-Level (ASGA): Replaces Softmax with learnable sparse gating, eliminating the strict sum-to-one constraint. (b) Meso-Level (GCQC): Evolves static target embeddings by fusing real-time context triggers. (c) Macro-Level (CGDF): A context-aware network that dynamically modulates fusion weights for heterogeneous views.}
    \vspace{-0.3cm}
    \label{fig:framework}
\end{figure*}

In this section, we present the proposed Gated Adaptive Progressive Network (GAP-Net), a unified framework designed to calibrate user intent via a multi-level gating philosophy. As illustrated in Figure~\ref{fig:framework}, GAP-Net establishes a "Triple Gating" architecture that filters noise across three granularities. The framework comprises three core modules: (1) Adaptive Sparse-Gated Attention for micro-level feature denoising; (2) Gated Cascading Query Calibration for meso-level intent evolution; and (3) Context-Gated Denoising Fusion for macro-level view modulation. We detail the design and implementation of each component in the following subsections.

\subsection{Problem Definition}
The objective of Click-Through Rate (CTR) prediction is to estimate the probability that a target user will engage with a candidate item within a specific context. Let $\mathcal{U}$ and $\mathcal{I}$ denote the sets of users and items, respectively. For a given instance comprising a user $u \in \mathcal{U}$, a candidate item $v^+ \in \mathcal{I}$, and a serving context $c$, the model input consists of static features and multi-granularity behavior sequences:
(1) Static Features: The user profile $\mathbf{x}_u$, candidate item attributes $\mathbf{x}_{v^+}$, and context features $\mathbf{x}_c$;
(2) Multi-Granularity Behavior Sequences: The user's historical interactions are partitioned into three temporal views: 
    Long-term behaviors $\mathbf{S}^{\mathrm{lt}}=[v_1, \ldots, v_{T_{\mathrm{lt}}}]$, 
    Short-term behaviors $\mathbf{S}^{\mathrm{st}}=[v_{T_{\mathrm{lt}}+1}, \ldots, v_{T_{\mathrm{lt}}+T_{\mathrm{st}}}]$, 
    and Real-time behaviors $\mathbf{S}^{\mathrm{rt}}=[v_{T_{\mathrm{lt}}+T_{\mathrm{st}}+1}, \ldots, v_T]$.

Here, $T = T_{\mathrm{lt}} + T_{\mathrm{st}} + T_{\mathrm{rt}}$ denotes the total sequence length. The typical magnitudes for these sequences are $T_{\mathrm{lt}} \sim 10^3\text{-}10^4$ (lifelong history), $T_{\mathrm{st}} \sim 10^2\text{-}10^3$ (recent interests), and $T_{\mathrm{rt}} \sim 10^0\text{-}10^2$ (current session triggers). 

All categorical and numerical features are mapped into dense embedding vectors $\mathbf{e}_u, \mathbf{e}_c, \mathbf{e}^{+} \in \mathbb{R}^d$. Similarly, each historical interaction item $v_t$ is embedded as $\mathbf{e}_{v_t} \in \mathbb{R}^d$, yielding the comprehensive behavior embedding matrices $\mathbf{E}_{\mathrm{lt}}$, $\mathbf{E}_{\mathrm{st}}$, and $\mathbf{E}_{\mathrm{rt}}$. The complete input representation for the CTR model is formalized as:
\begin{equation}
    \mathcal{X}=\left(\mathbf{e}_u, \mathbf{e}_c, \mathbf{e}^{+}, \mathbf{E}_{\mathrm{lt}}, \mathbf{E}_{\mathrm{st}}, \mathbf{E}_{\mathrm{rt}}\right).
\end{equation}

The model learns a mapping function $f: \mathcal{X} \mapsto \hat{y} \in [0,1]$, where $\hat{y}$ represents the predicted probability of a positive interaction (e.g., click or purchase). Given a training dataset $\mathcal{D}=\{(\mathcal{X}^{(i)}, y^{(i)})\}_{i=1}^N$ with binary labels $y^{(i)} \in \{0,1\}$, the parameters of $f(\cdot)$ are optimized by minimizing the binary cross-entropy loss:
\begin{equation}
    \mathcal{L}=-\frac{1}{N} \sum_{i=1}^N \left[ y^{(i)} \log \hat{y}^{(i)} + (1-y^{(i)}) \log (1-\hat{y}^{(i)}) \right],
\end{equation}
where $\hat{y}^{(i)} = f(\mathcal{X}^{(i)})$. 

Our work primarily focuses on designing the sequential encoding architecture within $f(\cdot)$. Specifically, we aim to model $\mathbf{S}^{\mathrm{lt}}$, $\mathbf{S}^{\mathrm{st}}$, and $\mathbf{S}^{\mathrm{rt}}$ in a disentangled, multi-granular, and context-aware manner, while seamlessly integrating them with static features for end-to-end CTR prediction.

\subsection{Adaptive Sparse-Gated Attention (ASGA)}
\label{subsec:asga}

Conventional Target Attention paradigms (e.g., DIN) typically employ linear projections for Query, Key, and Value, followed by Softmax normalization. However, this architecture faces two intrinsic limitations in modeling behaviors: (1) \textit{Representation Bottleneck}: Simple linear mappings lack the non-linearity required to capture high-order feature interactions between the target item and history; and (2) \textit{Noise Propagation}: Softmax normalization enforces a strict sum-to-one constraint, compelling the model to assign probability mass to noisy or irrelevant behaviors even when the current intent is semantically disconnected from past actions. This issue is analogous to the "Attention Sink" phenomenon observed in LLMs~\cite{xiao2023efficient}.

To address these challenges, we propose Adaptive Sparse-Gated Attention (ASGA). This module integrates Pre-Attention Feature Sifting (PAFS) to refine input embeddings and a Query-Guided Adaptive Output Gating (QAOG) mechanism to dynamically modulate the attention output based on the decision context.

\subsubsection{Pre-Attention Feature Sifting} 
Prior to computing interaction scores, it is essential to filter feature noise and enhance the representational capacity of the embeddings. Departing from standard linear projections, we implement a Gated Feed-Forward Network (SwiGLU-FFN) as a learnable feature sifter. 

Let $\mathbf{x} \in \mathbb{R}^d$ denote an input embedding (corresponding to either the target item $\mathbf{e}^+$ or a sequence item $\mathbf{e}_{v_t}$). We first project $\mathbf{x}$ into a higher-dimensional latent space $d'$ (defined as the next power of 2 relative to the hidden size) through two parallel layers: a gating path and an information path. We formally define the operation as follows:

\begin{equation}
\begin{aligned}
    \mathbf{h}_{\text{gate}} &= \text{Swish}(\mathbf{x}\mathbf{W}_g + \mathbf{b}_g) \\
    \mathbf{h}_{\text{up}} &= \mathbf{x}\mathbf{W}_u + \mathbf{b}_u
\end{aligned}
\end{equation}

where $\mathbf{W}_g, \mathbf{W}_u \in \mathbb{R}^{d \times d'}$ are the gate and up-projection matrices, respectively, and $\text{Swish}(z) = z \cdot \sigma(z)$ denotes the non-linear activation function. The feature filtering is performed via element-wise interaction, followed by a down-projection to restore the original embedding dimension:

\begin{equation}
    \text{PAFS}(\mathbf{x}) = (\mathbf{h}_{\text{gate}} \odot \mathbf{h}_{\text{up}})\mathbf{W}_d + \mathbf{b}_d
\end{equation}

where $\mathbf{W}_d \in \mathbb{R}^{d' \times d}$ is the down-projection matrix. Applying this transformation to our inputs yields:

\begin{equation}
    \tilde{\mathbf{e}}_t = \text{PAFS}(\mathbf{e}_t), \quad \tilde{\mathbf{E}}_s = \text{PAFS}(\mathbf{E}_s)
\end{equation}

This expansion-compression architecture functions as a learnable information bottleneck. The Swish gate ($\mathbf{h}_{\text{gate}}$) dynamically suppresses noisy signals within the expanded feature space, while the down-projection ($\mathbf{W}_d$) synthesizes the filtered features back into a compact representation, ensuring that only high-quality signals propagate to the subsequent attention mechanism.

\subsubsection{Query-Guided Adaptive Output Gating} 
Inspired by the Gated Attention mechanism in Qwen~\cite{qiu2025gated}, we adapt this concept for target attention. Unlike standard multi-head attention, we introduce a \textit{Query-Guided} gating mechanism. The fundamental premise is that the validity of the retrieved history is intrinsically dependent on the current target item (Query). If a user engages with an item contextually isolated from their history (e.g., accidental click), the model should autonomously suppress the historical context.

Specifically, we expand the Query projection to jointly learn the query representation and a relevance gate. For the $h$-th head:

\begin{equation}
    [\mathbf{Q}^{(h)}, \mathbf{G}_{\text{logit}}^{(h)}] = \text{Split}(\text{PAFS}(\mathbf{e}_t) \mathbf{W}_{Q}^{(h)})
\end{equation}

where $\mathbf{W}_{Q}^{(h)} \in \mathbb{R}^{d \times 2d_k}$. The projection is split along the last dimension to yield the query vector $\mathbf{Q}^{(h)} \in \mathbb{R}^{d_k}$ and the gating logit $\mathbf{G}_{\text{logit}}^{(h)} \in \mathbb{R}^{d_k}$. Keys ($\mathbf{K}$) and Values ($\mathbf{V}$) are projected normally from the sequence inputs:
\begin{equation}
    \quad \mathbf{K} = \tilde{\mathbf{E}}_s \mathbf{W}_K, \quad \mathbf{V} = \tilde{\mathbf{E}}_s \mathbf{W}_V
\end{equation}

We then modulate the standard Scaled Dot-Product Attention (SDPA) output $\mathbf{H}_{\text{att}}^{(h)}$ using the learned query gate:

\begin{equation}
    \begin{aligned}
         \mathbf{H}_{\text{att}}^{(h)} &= \text{Softmax}\left(\frac{\mathbf{Q}^{(h)}(\mathbf{K}^{(h)})^\top}{\sqrt{d_k}}\right)\mathbf{V}^{(h)} \\
        \mathbf{H}_{\text{final}}^{(h)} &= \mathbf{H}_{\text{att}}^{(h)} \odot \sigma(\mathbf{G}_{\text{logit}}^{(h)})
    \end{aligned}
\end{equation}

where $\sigma(\cdot)$ is the Sigmoid function. While Softmax determines the \textit{relative} importance distribution, the Sigmoid gate assesses the \textit{absolute} confidence of the query's need for context. If $\sigma(\mathbf{G}_{\text{logit}}^{(h)})$ approaches zero, the history is effectively ignored, eliminating the strict sum-to-one constraint and preventing noise propagation.

\subsection{Gated Cascading Query Calibration (GCQC)}
\label{subsec:gcqc}

Conventional sequential models typically operate under a \textit{Static Query Assumption}, treating the target embedding $\mathbf{e}^+$ as an immutable anchor. This creates a  \textit{semantic gap}: user intent toward the same target item is fluid and contingent on real-time context. Directly querying long-term history with a static target often retrieves contextually irrelevant behaviors.

To bridge this gap, we propose Gated Cascading Query Calibration (GCQC). Unlike static retrieval, GCQC adopts a \textit{Gated Hierarchy} strategy, employing a chain of Calibration Gating Units (CGU) to dynamically evolve the query vector.

\subsubsection{Hierarchical Sequence Partitioning}
Consistent with the Problem Definition, we stratify historical sequences (processed via PAFS) into three views:
\begin{itemize}[leftmargin=*]
    \item Real-Time View ($\tilde{\mathbf{E}}_{\mathrm{rt}}$): The most recent $T_{\mathrm{rt}}$ interactions, representing immediate triggers and impulse intent.
    \item Short-Term View ($\tilde{\mathbf{E}}_{\mathrm{st}}$): Recent window interactions, capturing temporary interests.
    \item Long-Term View ($\tilde{\mathbf{E}}_{\mathrm{lt}}$): Extensive history of stable preferences and habits.
\end{itemize}

\subsubsection{Gated Query Evolution}
Let $\text{ASGA}(\mathbf{Q}, \mathbf{S})$ denote the attention operation defined in \cref{subsec:asga}.The query evolves as follows:

\paragraph{Stage 1: Real-Time Context Injection}
The initial query is derived from the target item: $\mathbf{Q}_0 = \tilde{\mathbf{e}}_t$. We first query the Real-Time View $\tilde{\mathbf{E}}_{\mathrm{rt}}$. To fuse immediate context while preventing noisy drift, we introduce the first CGU:
\begin{equation}
\begin{aligned}
    \mathbf{H}_{rt} &= \text{ASGA}(\mathbf{Q}_0, \tilde{\mathbf{E}}_{\mathrm{rt}}) \\
    \mathbf{z}_1 &= \sigma([\mathbf{Q}_0; \mathbf{H}_{rt}]\mathbf{W}_{z1} + \mathbf{b}_{z1}) \\
    \mathbf{Q}_{rt} &= (1 - \mathbf{z}_1) \odot \mathbf{Q}_0 + \mathbf{z}_1 \odot \mathbf{H}_{rt}
\end{aligned}
\end{equation}

Here, $\mathbf{z}_1$ serves as an "Intent Update Gate".If real-time behaviors are relevant, $\mathbf{z}_1$ activates to inject context; otherwise, the gate closes to preserve original target semantics.

\paragraph{Stage 2: Short-Term Intent Rectification} 
Crucially, we use the Real-Time Calibrated Query $\mathbf{Q}_{rt}$—rather than the static target—to query the short-term history. This ensures retrieval is guided by the user's \textit{current} intent. A second CGU refines this representation:

\begin{equation}
    \mathbf{H}_{st} = \text{ASGA}(\mathbf{Q}_{rt}, \tilde{\mathbf{E}}_{\mathrm{st}}) 
\end{equation}

In this stage, $\mathbf{Q}_{rt}$ acts as a filter: only short-term behaviors that resonate with the real-time context are aggregated into $\mathbf{H}_{st}$. 

\paragraph{Stage 3: Context-Aware Long-Term Retrieval} 
As used in short-term modeling, $\mathbf{Q}_{rt}$ is also used to retrieve relevant memories from the extensive Long-Term View $\tilde{\mathbf{E}}_{\mathrm{lt}}$. Using a precise, context-aware query is vital for accurate retrieval from noisy long sequences:

\begin{equation}
    \mathbf{H}_{lt} = \text{ASGA}(\mathbf{Q}_{rt}, \tilde{\mathbf{E}}_{\mathrm{lt}})
\end{equation}

Through this gated cascade, GCQC effectively transforms the retrieval probability from the traditional $P(\text{History}|\text{Target})$ to a context-aware formulation $P(\text{History}|\text{Target}, \text{RealTime})$, ensuring query evolution is driven by high-confidence signals.

\subsection{Context-Gated Denoising Fusion (CGDF)}
\label{subsec:cgdf}

Existing frameworks typically adopt "hard concatenation" to merge temporal views, assuming all views are equally reliable. This inductive bias is flawed: view relevance is highly context-dependent (e.g., Long-Term dominates "repurchase", Real-Time dominates "impulse"). To address this, we propose Context-Gated Denoising Fusion (CGDF), which employs Gated Context Purification and Gated View Modulation.

\subsubsection{Gated Context Purification}

Let $\mathcal{V} = \{\mathbf{H}_{rt}, \mathbf{H}_{st}, \mathbf{H}_{lt}\}$ denote the calibrated representations from GCQC. To determine fusion weights, we construct a raw decision anchor $\mathbf{z}_{raw}$ by concatenating the refined target, context features, and view outputs:

\begin{equation}
    \mathbf{z}_{raw} = \text{Concat}(\tilde{\mathbf{e}}_t, \mathbf{e}_c, \mathbf{H}_{rt}, \mathbf{H}_{st}, \mathbf{H}_{lt})
\end{equation}

To filter noise (e.g., spurious correlations), we subject the anchor to a non-linear filtration using the SwiGLU-FFN architecture (identical to PAFS in Sec.~\ref{subsec:asga}):

\begin{equation}
    \mathbf{z}_{denoised} = \text{SwiGLU-FFN}(\mathbf{z}_{raw})
\end{equation}

This acts as a learnable filter, focusing the subsequent gating network on high-order interaction signals.

\subsubsection{Gated View Modulation}
The purified anchor $\mathbf{z}_{denoised}$ is passed through an MLP to project the context into a view-weighting space:

\begin{equation}
    \mathbf{h}_{gate} = \text{MLP}(\mathbf{z}_{denoised})
\end{equation}

Subsequently, we compute the adaptive fusion weights via a linear projection followed by Softmax normalization along the view dimension:

\begin{equation}
    \boldsymbol{\alpha} = \text{Softmax}(\mathbf{h}_{gate} \mathbf{W}_{logit})
\end{equation}

where $\boldsymbol{\alpha} = [\alpha_{rt}, \alpha_{st}, \alpha_{lt}] \in \mathbb{R}^3$ represents the learned importance distribution over the three temporal views, satisfying $\sum \alpha_k = 1$. The final fused representation $\mathbf{v}_{final}$ is obtained by the weighted aggregation (concatenation) of the expert views:

\begin{equation}
    \mathbf{v}_{final} = \text{Concat}(\alpha_{rt} \mathbf{H}_{rt}, \alpha_{st} \mathbf{H}_{st}, \alpha_{lt} \mathbf{H}_{lt})
\end{equation}

Here, scalar weights $\alpha_k$ are broadcast across embedding dimensions. This "Soft-Selection" mechanism empowers GAP-Net to dynamically suppress irrelevant temporal windows (e.g., $\alpha_{lt} \to 0$ during intent drift) while amplifying pertinent signals.

\section{Experiments}

In this section, we conduct extensive offline/online experiments to validate the effectiveness of GAP-Net on CTR tasks. The following \textbf{R}esearch \textbf{Q}uestions will be answered by analysis of the experimental results.
\begin{itemize}[leftmargin=*]
    \item \textbf{RQ1}: How does GAP-Net perform when compared with other state-of-the-art (SOTA) CTR models?
    \item \textbf{RQ2}: What is the influence on the performance of the core components in GAP-Net?
    \item \textbf{RQ3}: How do different components contribute to the effectiveness of GAP-Net?
    \item \textbf{RQ4}: How does GAP-Net perform in real industrial systems?
\end{itemize}

\subsection{Experimental Setting}

\subsubsection{Dataset} To validate the effectiveness of our method in recommendation systems, we construct a dataset XMart using real user interaction in our own scenario. Additionally, we evaluated our model on public recommendation dataset KuaiVideo\footnote{https://huggingface.co/datasets/reczoo/KuaiVideo\_XLong}. 

XMart. The XMart dataset which contains properties of users, items, user historically behaviors (including click, addcart, and purchase), is generated based on user logs collected from November 9th to 16th, 2025. User logs of the first 7 days in Nov. 2025 is used as the training data ,while reserve the last day for validation and test. Negative samples in the training dataset are set
to those impressed products but not purchased by users, and positive samples are purchased.

KuaiVideo. Derived from the Kuaishou Challenge presented at the China MM 2018 conference, this dataset is tailored for micro-video Click-Through Rate (CTR) prediction. It encapsulates a rich set of user-video interaction dynamics, recording specific behavior types including click'', like'', and follow'', alongside passive negative feedback (not click'' after thumbnail impression). For our experiments, we constructed a dense subset comprising 3,239,534 sequential interaction records sampled from 10,000 users. While absolute timestamps are anonymized, the relative temporal order of behaviors is preserved to facilitate sequential analysis. In this setup, positive samples are defined by explicit engagement behaviors (e.g., clicks), whereas impressions without subsequent clicks are treated as negative samples.

\subsubsection{Compared Methods}
To verify the effectiveness of the proposed method, we compare it with following methods:
\begin{itemize}[leftmargin=*]
    \item DIN~\cite{zhou2018deep}: It utilizes attention mechanism to activate relevant users’ behaviors with respect to corresponding targets and learns an adaptive representation vector for users’ interests.
    \item ETA~\cite{chen2021end}: It proposes an end-to-end target attention framework using Locality-Sensitive Hashing (SimHash). By retrieving top-k relevant behaviors via efficient Hamming distance calculation, it captures long-term user interests while satisfying strict inference time constraints.
    \item SDIM~\cite{cao2022sampling}: It introduces a hash sampling-based strategy to approximate the target attention distribution. By directly gathering behavior items that share the same hash signatures with the candidate item, it models long-term user interests with linear time complexity.
\end{itemize}

\subsubsection{Metrics}
To provide a comprehensive assessment of ranking efficacy, we employ three widely adopted metrics standard in industrial recommendation scenarios: AUC (Area Under ROC Curve) to measure the model's fundamental discriminative power, along with NDCG@K (Normalized Discounted Cumulative Gain) and MAP (Mean Average Precision) to evaluate the quality of the top-K ranking list.

\begin{table}[t]
    \centering
    \caption{Dataset Statistics}
    \label{tab:dataset}
    \begin{tabular}{ccccc}
       \toprule
         & users & items & inters. & avg inters. \\
       \midrule
        XMart & 8,678,328  &  25,033 & 1,463,105,174 & 168.59 \\
        KuaiVideo & 10,001 & 3,239,535 & 13,661,383 & 1366.14 \\
        \bottomrule
    \end{tabular}
\end{table}

Consistent with real-world serving systems, our evaluation is conducted on a \textit{per-request basis}. Specifically, for each user request $u$, the model scores and ranks a candidate set $\mathcal{I}_u$ composed of positive interactions $\mathcal{I}_u^+$ (clicked) and negative samples $\mathcal{I}_u^-$ (unclicked). The metrics are formalized as follows:

\begin{equation}
\begin{aligned}
    \text{AUC} &= \frac{1}{|\mathcal{D}|} \sum_{(i, j) \in \mathcal{D}} \mathbb{I}(\hat{y}_i > \hat{y}_j) \\
    \text{NDCG}@K &= \frac{1}{|\mathcal{U}|} \sum_{u \in \mathcal{U}} \left( \frac{\text{DCG}_u@K}{\text{IDCG}_u@K} \right) \\
    \text{MAP} &= \frac{1}{|\mathcal{U}|} \sum_{u \in \mathcal{U}} \left( \frac{1}{|\mathcal{I}_u^+|} \sum_{k=1}^{|\mathcal{I}_u|} P(k) \cdot r_k \right)
\end{aligned}
\end{equation}

where definitions are strictly detailed as:
\begin{itemize}[leftmargin=*]
    \item $\mathcal{D} = \{(i, j) | y_i=1, y_j=0\}$ denotes the set of all comparable positive-negative pairs across the dataset, and $\mathbb{I}(\cdot)$ is the indicator function.
    \item For top-K ranking, $r_k \in \{0, 1\}$ represents the ground-truth relevance at rank $k$, and $P(k)$ denotes the precision at cut-off $k$.
    \item $\text{DCG}_u@K = \sum_{k=1}^K \frac{2^{r_k} - 1}{\log_2(k+1)}$ accumulates the graded relevance with logarithmic decay, while $\text{IDCG}_u@K$ represents the score of the ideal ordering.
\end{itemize}

\subsubsection{Implementation Details} All models are implemented using Industrial-TensorFlow 1.15, Python 2.7, and trained on NVIDIA A100 GPU. For all models, we initialize the model parameters using the Xavier Initialization method~\cite{glorot2010understanding} and optimize the model with the Adam optimizer~\cite{kingma2014adam}, setting the learning rate to 0.001. The batch size is configured to 512.

\begin{table*}[t]
    \centering
    \caption{Comprehensive performance comparison on XMart (Click/Purchase) and KuaiVideos (Click).}
    \label{tab:overall_performance_full}
    \renewcommand{\arraystretch}{1.15}
    \setlength{\tabcolsep}{8pt} % Adjusted column spacing for better balance
    
    \begin{tabular}{l|ccc|ccc|ccc}
        \toprule
        \multirow{3}{*}{\textbf{Model}} & \multicolumn{6}{c|}{\textbf{XMart Dataset}} & \multicolumn{3}{c}{\textbf{KuaiVideos Dataset}} \\
        \cmidrule(lr){2-7} \cmidrule(lr){8-10}
         & \multicolumn{3}{c|}{\textbf{Click}} & \multicolumn{3}{c|}{\textbf{Purchase}} & \multicolumn{3}{c}{\textbf{Click}} \\
        \cmidrule(lr){2-4} \cmidrule(lr){5-7} \cmidrule(lr){8-10}
         & AUC & NDCG & MAP & AUC & NDCG & MAP & AUC & NDCG & MAP \\
        \midrule
        
        \textbf{DIN} & 0.6992 & 0.5401 & 0.3786 & 0.7587 & 0.5579 & 0.4142 & 0.6721 & 0.6999 & 0.3767 \\
        \textit{w/ GAP} & \textbf{0.7062} & \textbf{0.5453} & \textbf{0.3848} & \textbf{0.7661} & \textbf{0.5638} & \textbf{0.4213} & \textbf{0.6759} & \textbf{0.7057} & \textbf{0.3809} \\
        \midrule
        
        \textbf{ETA} & 0.6936 & 0.5362 & 0.3739 & 0.7547 & 0.5548 & 0.4104 & 0.6730 & 0.7021 & 0.3785 \\
        \textit{w/ GAP} & \textbf{0.7053} & \textbf{0.5446} & \textbf{0.3841} & \textbf{0.7647} & \textbf{0.5626} & \textbf{0.4198} & \textbf{0.6763} & \textbf{0.7062} & \textbf{0.3818} \\
        \midrule
        
        \textbf{SDIM} & 0.7034 & 0.5427 & 0.3816 & 0.7614 & 0.5599 & 0.4164 & 0.6738 & 0.7026 & 0.3792 \\
        \textit{w/ GAP} & \textbf{0.7056} & \textbf{0.5450} & \textbf{0.3844} & \textbf{0.7645} & \textbf{0.5630} & \textbf{0.4202} & \textbf{0.6792} & \textbf{0.7078} & \textbf{0.3834} \\
        \bottomrule
    \end{tabular}
\end{table*}

\subsection{Overall Performance (RQ1)} 
The comprehensive performance comparison on the XMart and KuaiVideos datasets is reported in Table~\ref{tab:overall_performance_full}. From the experimental results, we can draw three key observations regarding the effectiveness of GAP-Net:

\textbf{Universal Compatibility across Diverse Architectures.} A primary observation is that incorporating GAP-Net yields consistent and significant improvements across all baselines, regardless of their underlying modeling paradigms. Specifically, for DIN, which represents foundational Target-Attention models processing relatively short sequences, GAP-Net achieves an AUC lift of +1.00\% and +0.97\% on the XMart Click and Purchase tasks, respectively. More notably, for search-based models like ETA and SDIM, which are designed to handle ultra-long sequences, the performance gains are equally substantial (e.g., ETA + GAP improves XMart Click AUC by +1.69\% to reach 0.7053). This universality validates that the "Triple Gating" philosophy addresses fundamental bottlenecks common to both architectures: it not only fixes the "Inductive Bias Flaw" in standard attention (benefiting DIN) but also provides a dynamic calibration mechanism for long-sequence retrieval (benefiting ETA/SDIM), proving GAP-Net to be a robust, plug-and-play solution for sequential modeling.

\textbf{Enhanced Intent Resolution in High-Value Conversion Tasks.} Beyond basic Click-Through Rate (CTR) prediction, GAP-Net demonstrates superior efficacy on "high-value" conversion tasks, such as the Purchase task in XMart, which inherently requires a deeper understanding of user intent than simple clicks. As shown in Table~\ref{tab:overall_performance_full}, the relative improvements in Purchase prediction are particularly prominent. For instance, while DIN sees a steady increase in Click AUC, its gain in Purchase AUC is markedly strong, reaching 0.7661 (an improvement of +0.97\%). This phenomenon can be attributed to the \textit{Meso-Level Gated Cascading Query Calibration (GCQC)}. Conversion behaviors are often sparse and driven by highly specific, context-dependent intent. By progressively calibrating the query with real-time triggers, GAP-Net effectively filters out shallow "browsing noise" and accurately locks onto the strong purchase signals buried in the history, thereby delivering larger marginal gains on harder, intent-heavy tasks.

\textbf{Superior Ranking Stability via Noise Suppression.} In addition to binary classification metrics (AUC), the GAP-Net enhanced models exhibit remarkable gains in list-wise ranking metrics, specifically NDCG and MAP, across both datasets. For example, on the XMart dataset, DIN + GAP boosts NDCG from 0.5401 to 0.5453 (+0.96\%), and MAP from 0.3786 to 0.3848. Standard Softmax-based models suffer from the "Attention Sink" effect, where the model is forced to allocate probability mass to irrelevant noisy items, causing them to drift to the top of the recommendation list as "false positives." By implementing strict \textit{Adaptive Sparse-Gated Attention (ASGA)} at the micro-level, GAP-Net performs "soft rejection" on these low-confidence signals, effectively zeroing out noise. This cleans the decision boundary and ensures that the items ranked at the top-K positions are genuinely relevant to the user's calibrated intent, significantly improving the quality and stability of the final recommendation list.

\subsection{In-depth Analysis (RQ2 \& RQ3)}

\subsubsection{Ablation Study}
To verify the effectiveness of GAP-Net and quantify component contributions, we conducted a comprehensive ablation study on the XMart dataset, summarized in Table~\ref{tab:ablation}. First, replacing Softmax with ASGA confers a +0.35\% AUC uplift and improves NDCG to 0.5597, validating its efficacy in mitigating the ``Attention Sink'' effect via micro-level denoising. Second, incorporating GCQC brings a +0.28\% gain, confirming that real-time query calibration captures intent drift more effectively than static retrieval. Notably, CGDF contributes the largest individual increase of +0.44\%, demonstrating the clear superiority of dynamic view re-weighting over rigid hard concatenation. Ultimately, the full GAP-Net surpasses the baseline by +0.97\% in AUC and +1.05\% in NDCG. This substantial cumulative gain highlights the synergy of our architecture, proving that addressing noise simultaneously across micro, meso, and macro granularities is essential for robust modeling.

\begin{table}[t]
    \centering
    \caption{Ablation study of GAP-Net on the XMart dataset.}
    \label{tab:ablation}
    \renewcommand{\arraystretch}{1.1} 
    \setlength{\tabcolsep}{6pt} 
    \begin{tabular}{l|cccc}
        \toprule
        \multirow{2}{*}{Model Variants} & \multicolumn{3}{c}{XMart Dataset} \\
        \cmidrule(lr){2-4}
         & AUC & NDCG & MAP \\
        \midrule
        Baseline (No Gates)  & 0.7587 & 0.5579 & 0.4142 \\
        \midrule
        \quad + ASGA (Micro) & 0.7614 \small{(+0.35\%)} & 0.5597 & 0.4161 \\
        \quad + GCQC (Meso) & 0.7609 \small{(+0.28\%)} & 0.5593 & 0.4157 \\
        \quad + CGDF (Macro) & 0.7621 \small{(+0.44\%)} & 0.5605 & 0.4173 \\
        \midrule
        GAP-Net (Full) & \textbf{0.7661 \small{(+0.97\%)}} & \textbf{0.5638} & \textbf{0.4213} \\
        \bottomrule
    \end{tabular}
\end{table}

\subsubsection{Impact of Gating Strategy in ASGA}
To further elucidate the efficacy of the proposed micro-level denoising mechanism, we conduct a detailed comparative analysis in Table~\ref{tab:asga_compare}. First, we observe a counter-intuitive phenomenon: Naive Sigmoid (0.7563) actually underperforms the Standard Softmax baseline (0.7587). This reveals that merely removing the sum-to-one constraint is insufficient; without a competitive mechanism, unconstrained activation introduces optimization instability and fails to effectively distinguish signal from noise. In contrast, ASGA (0.7614) achieves significant improvements by implementing a controlled "soft rejection." The ablation results further validate our structural design: removing Pre-Attention Feature Sifting (w/o PAFS) or Query-Guided Gating (w/o QGG) leads to distinct performance drops (0.7614 → 0.7601 and 0.7597, respectively). This confirms that the synergy of feature-level sifting and intent-level gating is essential to effectively suppress the "Attention Sink" while maintaining representation stability, a balance that naive methods fail to achieve.

\begin{table}
    \centering
    \caption{Performance comparison of different attention activation strategies within the ASGA module.}
    \label{tab:asga_compare}
    \small
    \renewcommand{\arraystretch}{1.1}
    \setlength{\tabcolsep}{6pt}
    \begin{tabular}{l|ccc}
        \toprule
        \textbf{Attention Mechanism} & \textbf{AUC} & \textbf{NDCG} & \textbf{MAP} \\
        \midrule
        Standard Softmax (Baseline) & 0.7587 & 0.5579 & 0.4142 \\
        Naive Sigmoid (Direct Replacement) & 0.7563 & 0.5555 & 0.4112\\
        \midrule
        ASGA w/o PAFS & 0.7601 & 0.5593 & 0.4154 \\
        ASGA w/o QGG & 0.7597 & 0.5584 & 0.4147 \\
        \midrule
        \textbf{ASGA (Ours)} & \textbf{0.7614} & \textbf{0.5597} & \textbf{0.4161} \\
        \bottomrule
    \end{tabular}
\end{table}

\subsubsection{Impact of Gating Strategy in CGDF}
\begin{figure}
    \centering
    \includegraphics[width=\linewidth]{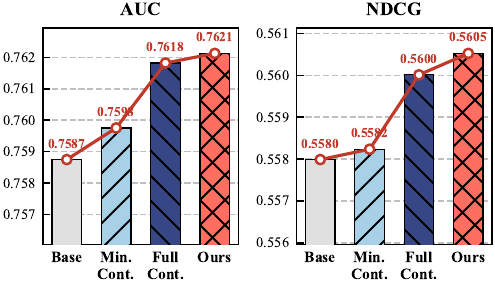}
    \caption{Impact of Gating Strategy in CGDF}
    \label{fig:cgdf_ablation}
    \vspace{-0.6cm}
\end{figure}

To investigate the optimal gating strategy in CGDF, we evaluate three context input variants against a static baseline: (1) \textit{Minimalist Context}, utilizing only target and sequence embeddings; (2) \textit{Full Context}, which naively concatenates all user profiles and context features; and (3) \textit{Purified Context}, our proposed method applying a denoising gate. The results in Figure~\ref{fig:cgdf_ablation} reveal that \textit{Minimalist Context} yields only a marginal AUC uplift (0.7587 → 0.7598), indicating that interaction embeddings alone lack the global perspective required for accurate routing. In contrast, \textit{Full Context} achieves a substantial jump to 0.7618, proving the value of incorporating rich side information; however, its potential is capped by the noise inherent in raw concatenation. Notably, \textit{Purified Context} surpasses all variants, reaching peak performance with 0.7621 AUC and 0.5605 NDCG. This distinct gain over the Full Context validates that simply expanding feature width is insufficient; the \textit{Gated Context Purification} is crucial for filtering low-level semantic noise, ensuring that the dynamic fusion mechanism is driven by high-fidelity signals rather than spurious correlations.

\subsection{Online A/B Test (RQ4)}
To rigorously evaluate the practical business value of our proposed model, we conducted a strictly controlled online A/B test on the Category List Page of Meituan Xiaoxiang Supermarket, a leading on-demand retail platform. The experiment spanned a 7-day window, involving live user traffic randomly bucketed into treatment and control groups. Compared to the highly optimized online baseline, our model achieved remarkable and consistent improvements across all core commercial metrics:

\begin{itemize}[leftmargin=*, topsep=0pt, partopsep=0pt]
    \item Gross Merchandise Value (GMV): We observed a robust +0.73\% lift in GMV. This substantial revenue gain indicates that our model not only promotes interactions but effectively identifies high-value potential needs, encouraging users to purchase items with higher unit prices or larger basket sizes.
    \item Conversion Rate (CVR): The model delivered a +0.57\% increase in CVR. This improvement in click-to-purchase efficiency validates that the items recommended by our model are genuinely aligned with users' immediate purchase intent, minimizing the gap between browsing and buying.
    \item Visit-to-Purchase Rate (V2P): Most notably, the Visit-to-Purchase Rate (defined as the ratio of paying users to total visiting users) saw a +0.33\% improvement. This metric serves as a direct proxy for the platform's overall conversion efficiency, demonstrating that our intent calibration mechanism successfully helps more hesitant browsers transition into paying customers.
\end{itemize}

\section{Conclusion}
In this paper, we propose GAP-Net, a unified Gated Adaptive Progressive Network that addresses the critical challenges of noise amplification and static intent assumptions in sequential recommendation. By establishing a systematic multi-level gating philosophy, GAP-Net integrates three core modules: Adaptive Sparse-Gated Attention (ASGA) to filter micro-level feature noise and mitigate the "attention sink" phenomenon; Gated Cascading Query Calibration (GCQC) to dynamically evolve user intent from real-time triggers to long-term memories; and Context-Gated Denoising Fusion (CGDF) to adaptively modulate heterogeneous temporal views based on decision context. This hierarchical architecture effectively bridges the semantic gap between static target items and dynamic user contexts. Extensive experimental results demonstrate that GAP-Net achieves state-of-the-art performance, exhibiting remarkable robustness against interaction noise and intent drift. These results validate the effectiveness of gating mechanisms in distilling complex user behaviors, providing a scalable and noise-resilient solution for next-generation recommendation systems.
% \begin{acks}

% \end{acks}

\bibliographystyle{ACM-Reference-Format}
\bibliography{sample-base}


\end{document}